\documentstyle[11pt,aaspp4,psfig]{article}   

\begin{document}

\slugcomment{To appear in The Astrophysical Journal.}

\title{Spin--Orbit Misalignment in Close Binaries with Two Compact
Objects}
\author{Vassiliki Kalogera}
\affil{Harvard-Smithsonian Center for Astrophysics, 60
Garden St., Cambridge, MA 02138; vkalogera@cfa.harvard.edu}

 \begin{abstract}
 Spin--orbit misalignment in coalescing compact binaries affects their
gravitational radiation waveforms. When the misalignment angles are large
($\gtrsim30^{\circ}$), the detection efficiency of the coalescence events
can decrease significantly if the misalignment effects are not modeled. In
this paper, we consider the formation of close compact binaries and
calculate the expected misalignment angles after the second core collapse
event. Depending on the progenitor parameters and the assumptions made
about supernova kicks, we find that 30\%--80\% of binaries containing a
black hole and a neutron star that coalesce within 10$^{10}$\,yr have
misalignment angles larger than 30$^\circ$ and a significant fraction of
them could remain undetected. The calculations allow us to place strong
constraints on the progenitors of such binaries and the kick magnitudes
required for their formation. We also discuss the formation of close
binaries with two black holes and the effect of non-isotropic kicks.
 \end{abstract}

\keywords{binaries: close --- stars: supernovae --- gravitation}

\section{INTRODUCTION}

The prototypical double neutron star system (NS--NS) is PSR B1913+16, the
first binary radio pulsar discovered 25 years ago (\cite{HT75}1975).  
Timing observations of the pulsar provided a remarkable confirmation of
general relativity with the measurement of several relativistic effects
including the orbital decay due to gravitational radiation
(\cite{TW82}1982). Since then three more such systems have been
discovered, including the recently detected candidate PSR J1811-1736
(\cite{L00}2000). The inspiral of such close NS--NS binaries continues
until the orbital separation becomes comparable to the NS radii and the
two stars merge on a dynamical timescale (\cite{RS92}1992;  
\cite{R96}1996). Such merger events are expected to occur not only in
NS--NS binaries but also in close black hole binaries, BH--NS and BH--BH.
These have not yet been directly observed, but their existence is
predicted by all theoretical models of binary evolution and compact object
formation (\cite{L97}1997; \cite{BB98}1998; \cite{PZ98}1998;
\cite{FWH99}1999). The final inspiral and coalescence of all 3 types of
close compact binaries are major sources of gravitational waves for the
laser-interferometer detectors currently under construction (LIGO, VIRGO,
GEO600; see \cite{T96}1996 for a review).

The detection of inspiral events relies on matched filtering techniques
(e.g., \cite{S91}1991), since the gravitational-wave signals are expected
to be weak relative to the various sources of detector noise. The
detection efficiency of binary coalescence depends on how extensive is the
database of ``search templates'', i.e., theoretically predicted inspiral
waveforms calculated for large ranges of parameters, such as masses,
detector location and orientation, and phase of the waves at coalescence.
The magnitude and orientation of the spins of the two compact objects
relative to the orbital angular momentum can also modify the inspiral
waveforms because of precession driven by general relativistic spin-orbit
and spin-spin couplings (e.g., \cite{A94}1994). \cite{A95}(1995) has
pointed out that the modulation of the gravitational-wave signal can be
large enough to decrease the detection efficiency significantly if
non-precessing waveforms are used in the search. The loss of detection
efficiency increases with (i) the mass contrast between binary components,
(ii) the spin magnitude of the more massive component, and (iii) the spin
misalignment from the angular momentum axis.  Even for maximally rotating
NS, it appears that an unmodulated family of templates would be sufficient
for the detection of NS--NS inspiral events. However, in the case of
BH--NS binaries (typical mass ratio $\sim 0.1$) with maximally rotating
BH, such a template family would be insufficient for more than 50\% of all
binary orientations if the spin tilt angle exceeds
30$^{\circ}$--40$^{\circ}$ and for {\em all} binary orientations if the
spin tilt angle exceeds 50--60$^{\circ}$ (\cite{A95}1995).

The ranges of parameters that the template database can realistically
cover is limited by the computational cost of computing a large number
of cross-correlations between templates and data. The range of spin
properties (magnitudes and orientations) is further limited by the
computational cost associated with calculations of precession-modulated
templates. Given these limitations it is important to examine (i)
whether it is necessary, based on astrophysical considerations, to
worry about the possibility of a significant misalignment, and (ii) how
extended a range of tilt angles should be covered by the precessing
templates.

In this paper we focus on the formation of close BH--NS binaries and we
determine the spin orientation of the BH at the time of their formation.
The evolution of BH--NS binary progenitors {\em prior} to the explosion
associated with the NS formation in the binary involves mass transfer
phases, which are expected to align the spins of both the BH and the NS
progenitor. Any spin tilt angle is therefore expected to be introduced by
the supernova (SN)  explosion that forms the NS. Mass loss alone cannot
misalign the two axes. However, there is now growing evidence that
asymmetric kicks are imparted to NS at birth (see \cite{FWH99}1999 and
references therein). In fact, as we show in \S\,3.1, the formation of a
coalescing BH--NS binary {\em requires} a significant NS kick.  Depending
on the kick magnitude and direction, the plane of the post-SN orbit can be
tilted relative to the pre-SN plane and hence the BH spin axis. We
consider a set of BH--NS progenitors (defined by the NS progenitor mass
and the pre-SN orbital separation) and calculate the theoretically
expected distributions of BH spin tilt angles for a number of isotropic
kick magnitude distributions. The results are most sensitive to the pre-SN
orbital separation and less sensitive to the assumed kick distribution. A
range of 30\%--80\% of BH-NS binaries are found to have tilt angles in
excess of 30$^{\circ}$. A discussion of the model assumptions is presented
in \S\,2.1 and the analysis of the asymmetric explosion is described in
\S\,2.2. Our results, along with a detailed parameter study are presented
in \S\,3. The effects of non-isotropic kicks are analyzed in \S\,4. The
significance of these results for the expected gravitational wave
detection efficiencies as well as our expectations for close BH--BH
binaries are discussed \S\,5.

\section{METHODS}

\subsection{Assumptions about Binary Progenitors}

Our current understanding of BH--NS binary formation (with a $\sim
10$\,M$_\odot$ BH) leads to an evolutionary history similar to that of
NS--NS binaries (\cite{vdH76}1976). The difference is that one of the
two stars, normally the primary, is massive enough to collapse into a
BH. When the primary evolves away from the main sequence and expands,
it fills its Roche lobe, transfers mass to its companion and eventually
its core collapses to form a BH. The system becomes a high-mass X-ray
binary (such as Cyg X-1) until the secondary evolves and expands enough
to fill its Roche lobe. At this point the mass transfer is almost
certainly unstable and the binary goes through a common-envelope phase.
The result is a much tighter binary containing the BH and the helium
core of the secondary. The last stage involves the supernova explosion
and collapse of the helium star into a NS and a BH--NS binary is
formed.  Variations on this main evolutionary path are possible but
detailed calculations show that their relative fraction is negligible
(\cite{FWH99}1999).

Here we study the formation of BH--NS coalescing binaries, but we consider
only their {\em immediate} progenitors, i.e., systems just before the
second core collapse in the binary. There are two reasons for this
approach. Modeling the complete evolutionary history of BH--NS binaries is
rather uncertain because it involves a long sequence of poorly understood
phases, such as non-conservative, stable or unstable mass transfer, wind
mass loss from hydrogen- and helium-rich stars, and BH formation with mass
loss and possibly kicks. A wide range of assumptions are necessary to
describe the evolution and they can affect the results in a complicated
way. Instead, we study BH--NS formation for specific pre-SN parameters and
we obtain significant constraints on these parameters.  This approach
frees our calculations from a large set of uncertain assumptions
concerning earlier stages. More importantly, it allows us to understand
the physical origin of the dependence of our results on the few input
parameters, and to examine the robustness of the results.

As described above, we expect that the BH progenitor was the more
massive of the two binary members and that the BH formed first in the
system. It is in principle possible, through mass transfer episodes,
that a mass ratio inversion occurred during the evolutionary history of
the system. The result in this case would be that the collapse order
was reversed and that the NS formed before the BH.  The immediate
BH--NS binary progenitor would then consist of a NS and a
non-degenerate star massive enough to collapse into a BH. However,
detailed evolutionary calculations covering a wide range of model
parameters indicate that the fraction of primordial binaries
experiencing such a mass ratio inversion is negligible (Fryer 1999,
private communication).

We assume that the orbit before the SN explosion is circular. This is
well justified since it is thought that the binary experienced a phase
of unstable mass transfer and common-envelope evolution.  In principle,
one could doubt that this phase occurred, but we will show in \S\,3.1
that it is actually {\em necessary\/} for the formation of {\em
coalescing} BH--NS binaries. In the absence of common-envelope
evolution and the resulting orbital contraction, the pre-SN separation
would very large ($\sim 10^3$\,R$_\odot$), so that the massive NS
progenitor would never fill its Roche lobe. However, we will show that
such wide binaries cannot be the progenitors of BH-NS systems tight
enough to coalesce within a Hubble time. Given the highly dissipative
nature of the common envelope phase, it seems inevitable that the
binary orbit will be circularized.

Another inevitable outcome of the mass transfer phases occurring prior
to NS formation is the alignment of the spin axes of the BH and NS
progenitors with the orbital angular momentum. It is this expectation
of alignment that allows us to calculate the spin tilt angle
distribution of BH--NS binaries, since we can identify the tilt angle,
$\omega$, of the post-SN orbital plane relative to the pre-SN plane
with the misalignment angle of the BH spin relative to the orbital
angular momentum in the BH--NS system (post-SN). We note that it is the
BH spin orientation that has been found to be more important in
modifying BH--NS inspiral waveforms (\cite{A94}1994).

The results presented in \S\,3 are obtained under the assumption
that kicks imparted to NS are distributed isotropically. In the absence of
a clear picture of the physical origin of kicks this assumption seems
reasonable. However, for reasons of completeness, we also consider
(section 4) the case of non-isotropic kicks that are directed
preferentially perpendicular to or in the pre-SN orbital plane.

Finally, in the orbital dynamics analysis presented here we ignore the
impact of the supernova shell on the NS companion. This is well
justified because the NS companion is a BH with a negligible cross
section (even in the case of a non-degenerate 1\,M$_\odot$ companion,
the impact becomes important {\em only\/} if the pre-SN orbital
separation $\lesssim 3$\,R$_\odot$, see e.g., \cite{R92}1992 and
\cite{K96}1996).

In our choice of the parameter values for our standard case, a
10\,M$_\odot$ BH, a 4\,M$_\odot$ NS progenitor, and a 10\,R$_\odot$ pre-SN
orbital separation, we are motivated by (i) the range of covered by BH
mass measurements in soft X-ray transients (\cite{C98}1998), (ii) the
general picture of BH--NS formation, according to which their immediate
progenitors are the end products of common-envelope evolution, and hence
the BH companions have lost their hydrogen-rich envelopes and the orbits
are tight ($\sim 10$\,R$_\odot$), and (iii) calculations of the evolution
of helium stars with wind mass loss that lead to final masses of
$3-4$\,M$_\odot$ (\cite{WLW95}1995) and the results of core-collapse
simulations suggesting that helium stars more massive than
$7-10$\,M$_\odot$ collapse into black hole instead of a neutron star
(\cite{F99}1999).

\subsection{Post-Supernova Tilt Angles}

We consider a binary consisting of a BH of mass $M_{\rm BH}$ and a
non-degenerate star (the NS progenitor) of mass $M_0$
in a circular orbit of separation $A_0$. We assume that $M_0$ explodes
instantaneously, i.e., on a timescale shorter than the orbital period,
leaving a NS remnant of mass $M_{\rm NS}$, and that a kick of magnitude
$V_k$ is imparted to the remnant. The post-SN characteristics, orbital
separation $A$, eccentricity, $e$, and tilt of the orbital plane
$\omega$, can be derived based on conservation laws and the geometry of
the system (see e.g., \cite{H83}1983; \cite{BP95}1995; \cite{K96}1996).

From energy conservation we obtain
 \begin{equation}
 \alpha~\equiv ~\frac{A}{A_0}~=~\frac{\beta}{2\beta-u_k^2\,\sin^2\theta
-(u_k\cos\theta+1)^2},
 \end{equation}
 where $u_k\equiv V_k/V_r$ is the kick magnitude in units of the pre-SN
relative orbital velocity $V_r$
 \begin{displaymath}
 V_r~=~\left(G\,\frac{M_{\rm BH}+M_0}{A_0}\right)^{1/2},
 \end{displaymath}
 $\beta$ is the ratio of the total mass after and before the explosion,
 \begin{displaymath}
 \beta~=~\frac{M_{\rm BH}+M_{\rm NS}}{M_{\rm BH}+M_0},
 \end{displaymath}
 and the angles $\theta$ and $\phi$ define the direction of the kick:
$\theta$ is the polar angle from the pre-SN orbital velocity vector of the
exploding star ($m$) and ranges from $0-\pi$ (at $\theta=0, \vec{V_k}$ and
$\vec{V_r}$ are aligned); $\phi$ is the azimuthal angle in the plane
perpendicular to $\vec{V_r}$ (i.e., $\theta=\pi /2$) and ranges from
$0-2\pi$ (at $\theta=\pi /2$ and $\phi=0$ or $\phi=\pi$, the kick
component points along or opposite of the angular momentum axis of the
pre-SN orbital plane, respectively; Figure 1).

From orbital angular momentum conservation we obtain for the
post-SN eccentricity
 \begin{equation}
 1-e^2~=~\frac{1}{\beta^2}\left[u_k^2\,\sin^2\theta\,\cos^2\phi+
(u_k\cos\theta+1)^2\right]\,\left[2\beta-u_k^2\,\sin^2\theta
-(u_k\cos\theta+1)^2\right].
 \end{equation}

The tilt angle $\omega$ between the orbital planes before and after the
explosion is equal to the angle between the vector $\vec{V_r}$ and the
projection $\vec{V_p}$ of $\vec{V_r}+\vec{V_k}$ onto the plane defined by
$\phi=\pi/2$, (which contains $\vec{V_r}$ and is perpendicular to the
binary axis). Note that the intersection of the pre- and post-SN planes
lies along the binary axis. Evaluating the dot product
$\vec{V_r}\cdot\vec{V_p}$ and using equation (1) we obtain (see also
\cite{K96}1996)
 \begin{eqnarray}
\cos\omega
& = &
\left(u_k\,\cos\theta\,+\,1\right)\,\left[u_k^2\,\sin^2\theta\,
  \cos^2\phi+(u_k\,\cos\theta\,+\,1)^2\right]^{-1/2} \nonumber \\
 & = &
\left(u_k\,\cos\theta\,+\,1\right)\,\left(2\beta-\frac{\beta}{\alpha}-
u_k^2\,\sin^2\theta\,\sin^2\phi\right)^{-1/2}
 \end{eqnarray}

For a given set of pre-SN binary parameters (masses and orbital
separation) and a fixed kick magnitude, there are only two free
parameters in the problem: the kick direction angles $\theta$ and
$\phi$. Therefore only two of the post-SN characteristics are truly
independent parameters. An assumed distribution ${\cal F}(\theta,\phi)$
for the two kick angles can be transformed into a probability
distribution for any two post-SN parameters. Here we are interested in
the tilt of the orbital plane so we calculate the Jacobian
transformation
 \begin{equation}
 \cal{F}^\prime\left(\alpha,\omega\right)~=~\cal{F}\left(\theta,\phi
\right)\,J\left(\frac{\theta,\phi}{\alpha,\omega}\right)~=~
\cal{F}\left(\theta,\phi\right)\,\left(\frac{\partial\theta}{\partial\alpha}\,
\frac{\partial\phi}{\partial\omega}\,-\,\frac{\partial\theta}{\partial\omega}
\,\frac{\partial\phi}{\partial\alpha}\right),
 \end{equation}
 where ${\cal F}^\prime\left(\alpha,\omega\right)$ is the probability
distribution for $\alpha$ (eq.\ [1]) and the tilt angle $\omega$. 

 For an isotropic kick distribution we have
 \begin{equation}
{\cal F}\left(\theta,\phi\right)~=~\frac{\sin\theta}{2}\,\frac{1}{2\pi}.
 \end{equation}
 Inverting equations (1) and (3) we get
 \begin{eqnarray}
 \cos\theta & = &
\frac{1}{2u_k}\,\left(2\beta-\frac{\beta}{\alpha}-u_k^2-1\right),
\nonumber \\
 \sin^2\phi & = & 4\left[4u_k-\left(2\beta-\beta/\alpha-u_k^2-1\right)
^2\right]^{-1}\left[2\beta-
\frac{\beta}{\alpha}-\left(\frac{2\beta-\beta/\alpha-u_k^2+1}{2\cos\omega}
\right)^2\right].
 \end{eqnarray}
 Equations (4) and (5) then give
 \begin{equation}
 {\cal F}^\prime\left(\alpha,\omega\right)=\frac{\beta}{2\pi
u_k\alpha^2}\left[\frac{4(2\beta-\beta/\alpha)}{(2\beta-\beta/\alpha-
u_k^2+1)^2}\,\cos^2\omega-1\right]^{-1/2}\left(1-\cos^2\omega\right)^{-1/2}
\vert\cos^{-1}\omega\vert.
 \end{equation}
 Finally, to obtain the probability distribution
$\cal{F}_{\omega}\left(\omega\right)$, we have to integrate
$\cal{F}^\prime\left(\alpha,\omega\right)$ over $\alpha$ with
appropriate bounds. These bounds are dictated by two requirements: (i)
the SN explosion does not disrupt the binary and (ii) the post-SN
system is a {\em coalescing binary}, i.e.  it will coalesce within a
Hubble time ($\sim 10^{10}$\,yr). Note that the integral of
$\cal{F}_{\omega}\left(\omega\right)$ over $\omega$ is not equal to unity
but instead equal to the fraction of BH--NS binaries that satisfy the
above two constraints. 

\section{Results}

\subsection{Limits on the Spin Tilt Angle}

Before we go on with the calculation of spin tilt angle distributions, we
start by deriving limits on the tilt angle $\omega$, given a set of pre-SN
parameters ($M_{\rm BH}$, $M_0$, and $A_0$). We derive these limits using
equation (7) and imposing the obvious requirements that
 \begin{equation}
 -1~\leq~\cos\theta~\leq~1 \hspace{2.cm} \sin^2\phi~\leq~1, 
 \end{equation}
 and that the post-SN system is bound and will coalesce within
$10^{10}$\,yr. The latter two constraints translate into limits on the
orbital separation $A$ for a given post-SN eccentricity $e$ (eq.\ [2]).
Keeping the system bound requires (\cite{FvdH75}1975)
 \begin{equation}
 \frac{1}{1+e}~<~\frac{A}{A_0}~<~\frac{1}{1-e}. 
 \end{equation}
 The condition that the coalescence time be shorter than $10^{10}$\,yr
translates into an upper limit on $A$ for a given $e$. We calculate
this limit using expressions derived by \cite{JS92}(1992) and we plot
it in Figure 2, for BH--NS systems and NS--NS systems, for comparison.
In this and all subsequent Figures we have adopted $M_{\rm
NS}=1.4$\,M$_\odot$, since measured NS masses are all consistent with a
narrow range around this value (see \cite{TC98}1998).

The limits on the tilt angle $\omega$ are shown in
Figure 3 as a function of the isotropic kick magnitude, $V_k$, and for
different values of $M_{\rm BH}$, $M_0$, and $A_0$.  It is evident that
the requirements of equations (8) and (9), and that coalescence occurs
within $10^{10}$\,yr, lead to constraints not only on the tilt angle
but also on the kick magnitude and the pre-SN separation. The lower
limit on kick magnitude arises from the requirement that post-SN
systems should coalesce within $10^{10}$\,yr. A minimum kick is
necessary to overcome the orbital expansion resulting from the mass
loss at NS formation. If there is no kick, or if the kick magnitude is
too low, post-SN systems are too wide and have coalescence times longer
than $10^{10}$\,yr. The upper limit on kick magnitude arises from both
requirements that post-SN systems are bound {\em and} coalesce within
$10^{10}$\,yr. If the kick is too large, then most systems get
disrupted, while those that remain bound have wide orbits and long
coalescence times. Previous analyses of the effect of kicks on orbital
dynamics (e.g.,~\cite{H83}1983; \cite{K96}1996) have shown that
formation of post-SN systems that satisfy constraints similar to those
considered here require that (i) kick magnitudes be of the order of the
pre-SN orbital velocity, $\sim 500$\,km\,s$^{-1}$ for our standard
case, and (ii) kicks be directed close to the orbital plane and
opposite to the velocity vector of the exploding star
($\theta\sim\pi$).  For a given kick magnitude close to the minimum
value required, there is a limit on how large the kick component
perpendicular to the pre-SN orbital plane can be. This component is
responsible for the tilt of the plane, and hence an upper limit to the
plane tilt angle $\omega$ exists. As the kick magnitude increases and
approaches the pre-SN orbital velocity, the range of allowed tilt
angles becomes wider. As the kick magnitude increases further and
becomes larger than the pre-SN orbital velocity, a lower limit on the
tilt angle appears because there is always some excess kick component
perpendicular to the pre-SN orbital plane.  This qualitative behavior
of the allowed tilt angles with an increasing kick magnitude is quite
robust and independent of the assumed pre-SN binary parameters. The
derived values of the minimum and maximum kick magnitudes depend of
course on the values of $M_{\rm BH}$, $M_0$, and $A_0$ (see Figure 3).
For our standard case, $M_{\rm BH}=10$\,M$_\odot$, $M_0=4$\,M$_\odot$,
and $A_0=10$\,M$_\odot$, coalescing BH--NS binaries can be formed only
if $50$\,km\,s$^{-1} < V_k < 1000$\,km\,s$^{-1}$.

From the discussion above it becomes evident that the
range of required kick magnitudes is determined by 
 \begin{equation}
 V_k~\sim~V_r~=~\left(G\,\frac{M_{\rm BH}+M_0}{A_0}\right)^{1/2}. 
 \end{equation} 
 Hence the required kick values decrease with increasing $A_0$, i.e.,
for wider and less bound pre-SN binaries. The range of allowed kicks
also becomes narrower with increasing $A_0$ since the pre-SN binary is
less bound and is much easier to disrupt once the kick exceeds the
orbital velocity. This dependence on $A_0$ allows us to derive a strong
{\em upper limit} on its value. For the case of $M_{\rm
BH}=10$\,M$_\odot$, it must be $A_0~<~300$\,R$_\odot$. Pre-SN binaries
in wider orbits are so loosely bound that the kicks that would allow
them to remain bound after the explosion are {\em too low\/} to decrease
the post-SN orbital separation enough, to make the coalescence time
$<10^{10}$\,yr. Therefore, there is no kick magnitude that
allows such wide systems to be both bound and coalescing after the SN.

This upper limit on $A_0$ is important because it strongly constrains
the nature of the NS progenitor. We mentioned above that the NS
progenitor is expected on evolutionary grounds to be a helium star, the
core of the hydrogen-rich NS progenitor exposed at the end of a
common-envelope phase. The derived upper limit on $A_0$ supports this
expectation. Had the NS progenitor just before the SN been a massive,
hydrogen-rich star, the orbital separation $A_0$ would have to be $\sim
10^3$\,R$_\odot$ (e.g., \cite{S92}1992), to accommodate the radial
expansion of the evolved star. For any mass of the NS progenitor
appropriate for a hydrogen-rich star ($10-25$\,M$_\odot$, see
\cite{FK99}1999), such a configuration can be safely excluded, since
the required kick magnitude range vanishes. We conclude, therefore,
that immediate BH--NS binary progenitors must contain a BH and a
helium-star in orbits with separations $\lesssim 300$\,R$_\odot$. Such
systems can be formed only through a common-envelope phase.  Exposure
of the helium core through strong mass loss (e.g., \cite{S92}1992;
\cite{WL99}1999) can also be excluded for BH--NS progenitors, since the
binary orbit expands during such a phase, instead of contracting.
Although the upper limit on $A_0$ is $\sim 300$\,R$_\odot$, for
common-envelope (CE) evolution, the {\em typical} values of the post-CE
orbital separations are much lower, $\sim 10$\,R$_\odot$ (e.g.,
\cite{KW98}1998).

In agreement with equation (10), an increase in $M_0$ favors higher
kick magnitudes (Figure 3). Here we consider a range of NS progenitor
masses appropriate for helium stars. The minimum helium-star mass for
NS formation has been estimated to be $2-3$\,M$_\odot$ (e.g.,
\cite{H86}1986) and, based on current core-collapse
calculations (\cite{F99}1999), the upper limit lies probably in the
range $7-10$\,M$_\odot$. We plot our results for two values of $M_0$,
4\,M$_\odot$ and 10\,M$_\odot$. It is evident that the dependence of
our results on $M_0$ in such a small range is very weak (Figure 3).

The dependence of the kick and tilt angle limits on the BH mass also
follows equation (10), as expected. Here we consider three different
values here, $M_{\rm BH}=5, 10,~{\rm and}~20$\,M$_\odot$. We note that,
given our present understanding of massive star evolution with mass
loss, BH masses in excess of $\simeq 20$\,M$_\odot$ are not favored
(\cite{FK99}1999). 

In Table 1 we summarize the various sets of pre-SN parameters we
consider here, the corresponding pre-SN orbital velocities, and the
limits imposed on the isotropic kick magnitude.

\subsection{Spin Tilt Angle Distributions} 

\subsubsection{Fixed Kick Magnitude}

For given values of $M_{\rm BH}$, $M_0$, $A_0$, and $V_k$, we calculate
the probability distributions of spin tilt angle $\omega$ for coalescing
BH--NS binaries, as described in \S\,2.2 (integrating Eq.\ [7]
numerically). The results, for our standard case ($M_{\rm
BH}=10$\,M$_\odot$, $M_0=4$\,M$_\odot$, and $A_0=10$\,R$_\odot$), are
shown in Figure 4 (top) for different values of the kick magnitude. Note
that in this plot the integral of each distribution is not equal to unity
but instead equal to the fraction of post-SN systems that remain bound and
will coalesce within 10$^{10}$\,yr.  We also plot the normalized to unity
cumulative angle distributions, i.e., the fraction of systems with tilt
angles smaller than a given value $\omega$, in Figure 4. The behavior
described in \S\,3.1 is even more clearly seen here. For the case shown in
Figure 4, the pre-SN relative orbital velocity is $V_r \simeq
520$\,km\,s$^{-1}$.  For low kick magnitudes, the tilt angles are
restricted to small values. As the kick magnitude increases the allowed
range of angles widens and the peak of the angle distribution within this
range shifts to its high end. For kicks comparable to or slightly higher
than $V_r$, tilt angles in the full range from alignment ($\omega\sim
0^{\circ}$) to anti-alignment ($\omega\sim 180^{\circ}$) with the orbital
angular momentum axis are possible. For even higher kicks the tilts are
restricted to only high values ($>90^{\circ}$, i.e., retrograde post-SN
orbits).

\subsubsection{Kick Magnitude Distributions} 

We can take one step further and examine the distribution of BH--NS tilt
angles not just for a given kick magnitude but for an assumed distribution
of kick magnitudes. The calculation involves the convolution of the
previously derived probability distributions of $\omega$ with a kick
magnitude distribution,
 \begin{equation}
 {\cal T}(\omega; M_{\rm BH}, M_0, A_0)~=~\int~{\cal
F}_{\omega}\left(\omega; V_k, M_{\rm BH}, M_0, A_0\right)~F_K(V_k)~dV_k.
 \end{equation}

Since the physical origin of NS kicks is not well understood at present,
it is not possible to predict theoretically their magnitude distribution.
Instead, there have been several attempts to derive a kick distribution
based on observational constraints, primarily from transverse radio pulsar
velocity measurements (e.g., \cite{HP97}1997; \cite{CC98}1998), but also
using other populations (e.g., \cite{FBB98}1998; \cite{KKK98}1998). A
Maxwellian form (Gaussian kick components in all three directions) has
often been assumed and the velocity dispersion, $\sigma$, can then be
fitted to observations. Different $\sigma$ values have been derived
depending on considerations of selection effects and measurement errors
for the various NS populations. Overall, a consensus seems to have formed,
placing the average kick magnitude in the range $100-500$\,km\,s$^{-1}$.
Here we calculate the final tilt angle distributions for two Maxwellian
distributions with $\sigma=100, 200,~{\rm and}~400$\,km\,s$^{-1}$, 
and for one extreme case of a flat distribution in the
range $0-1500$\,km\,s$^{-1}$. The results are shown in Figure 5
(distribution functions and normalized cumulative distributions) for our
standard case. The dependence of the resulting distribution on the average
kick magnitude shows the expected trend, i.e., the higher the average
kick, the smaller the fraction of BH--NS binaries with small tilt angles
(e.g., $\omega<30^{\circ}$; see Figure 5). The results appear to be
remarkably robust in the two cases of a Maxwellian with a relatively high
$\sigma$ and a flat distribution. The origin of this robustness is that
the shape of the tilt distribution is not determined by the overall shape
of the kick magnitude distribution, but instead by the shape of the
distribution (or fraction of kicks) within the range of magnitudes
required for BH--NS formation, given the assumed pre-SN parameters.

In Figure 6, we show the dependence of the final tilt distributions for
different sets of pre-SN parameters and for a Maxwellian kick distribution
($\sigma=200$\,km\,s$^{-1}$). It is evident that the fraction of
coalescing BH--NS binaries with tilt angles higher than $30^{\circ}$
increases as the immediate progenitors becomes more loosely bound.

\section{NON--ISOTROPIC KICKS} 

So far we have assumed that NS kicks are directed isotropically. However,
it is possible that certain directions are favored because of the unknown
details of the physical mechanism responsible for the kick. In what
follows we examine two cases where kicks are preferentially directed
either (i) perpendicular to the pre-SN orbital plane along two cones with
axes parallel to the pre-SN orbital angular momentum axis and with an
assumed opening angle $\theta_p$ (i.e., {\em polar} kicks), or (ii) close
to the pre-SN orbital plane or else perpendicular to the angular momentum
axis in a ``fan'' shape with an assumed half--opening angle $\theta_p$
(i.e., {\em planar} kicks). The angle $\theta_p$ can vary between
0$^{\circ}$ and 90$^{\circ}$.

These two cases of non-isotropic kicks translate into certain constraints
imposed on the two angles $\theta$ and $\phi$ that determine the kick
direction in the reference frame defined in section \S\,2.1. We derive
these constraints using another reference frame, in which the polar
angle $\theta^{\prime}$ is defined with respect to the pre-SN angular
momentum axis (the axis out of the page, towards the reader in Figure 1)
instead of $\vec{V_r}$. The two frames are connected by the condition
 \begin{equation}
 \cos\theta^{\prime} = \sin\theta\cos\phi. 
 \end{equation}

For the two cases of anisotropicity we consider here, the constraints
imposed on $\theta^{\prime}$ are
 \begin{equation}
 \cos\theta_p \leq \cos\theta^{\prime} \leq 1 \hspace{1.cm} {\rm and}
\hspace{1.cm} -1 \leq \cos\theta^{\prime} \leq -\cos\theta_p,  
 \end{equation}
 for polar kicks and 
 \begin{equation}
 -\sin\theta_p \leq \cos\theta^{\prime} \leq \sin\theta_p, 
 \end{equation} 
 for planar kicks. These translate into constraints on $\theta$ and
$\phi$: 
 \begin{equation}
 \cos\theta_p \leq \vert\,\sin\theta\,\cos\phi\,\vert \leq 1
 \end{equation}
 and 
 \begin{equation}
 0 \leq \vert\,\sin\theta\,\cos\phi\,\vert \leq \sin\theta_p, 
 \end{equation} 
 for polar and planar kicks, respectively. The latter two constraints
substitute those given in equation (8) for isotropic kicks. 

Using equations (15) and (16)  we can calculate the limits on the tilt
angle $\omega$ based on the analysis presented in \S\,2.1. The results are
shown in Figure 7 for the case of $M_{\rm BH}=10$\,M$_\odot$,
$M_0=4$\,M$_\odot$, $A_0=10$\,R$_\odot$, and for angles $\theta_p$ varying
from 90$^{\circ}$ to 10$^{\circ}$. Note that $\theta_p=90^{\circ}$
corresponds to the case of isotropic kicks.

In the case of polar kicks, i.e., kicks constrained in two cones along the
orbital angular momentum axis (top panel in Figure 7), the effect of
anisotropicity is more prominent than in the case of planar kicks, i.e.,
kicks constrained to be within an angle of the pre-SN orbital plane
(bottom panel in Figure 7). This is an indirect demonstration of the fact
that, even when all kick directions are allowed, the requirement that
post-SN systems are bound in tight orbits acts as a filter and planar
kicks are preferred (e.g., \cite{H83}1983; for binary compact objects, see
\cite{WKK00}2000). The top panel of Figure 7 indicates that, as the
opening angle of the cones decreases, the range of allowed tilt angles and
kick magnitudes shrinks. For the specific choice of masses shown, no
coalescing binaries can form if $\theta_p\leq 30^{\circ}$. On the other
hand, in the bottom panel, the limits are altered significantly from those
in the isotropic case only for $\theta_p\lesssim 30^{\circ}$, when the
``fan--shaped'' region closes into the pre-SN orbital plane.

The effects of non--isotropic kicks on the range of allowed tilt angles as
a function of the kick magnitude (see Figure 7) can be understood based on
two considerations: (i) the kick component out of the pre-SN orbital plane
is primarily responsible for the tilt, and (ii) bound post-SN systems in
tight orbits can be formed only when the pre-SN orbital velocity of the
exploding star and the kick component opposite the orbital motion are
roughly comparable in magnitude. In the case of polar kicks with low
magnitudes, the magnitude of the kick component in the orbital plane
becomes restricted.  Therefore bound systems are formed with higher and
higher tilt angles as the kick anisotropicity away from the orbital plane
becomes stronger. For moderate magnitudes, very low tilt angles are not
allowed for the same reason, but very high tilts are also disfavored
because the binaries become either too wide or get disrupted altogether,
especially for high total kick magnitudes. As we already mentioned, the
effects are less dramatic in the case of planar kicks. Low kick magnitudes
tend to favor close binaries with small tilt angles. On the other hand,
for large kicks directed within a very small angle from the orbital plane
(e.g., 10$^{\circ}$), the kick component in the plane tends to be too
large and systems again become too wide or get disrupted.

In Figure 8 we show the normalized cumulative distributions of tilt angles
already convolved with a kick magnitude distribution (Maxwellian with
$\sigma=200$\,km\,s$^{-1}$). As expected based on our understanding, for
kicks increasingly restricted in directions away from the pre-SN
orbital plane, the fraction of coalescing BH--NS binaries with small tilt
angles (for example, $< 30^{\circ}$) decreases. For kicks increasingly
restricted to lie close to the plane, the same fraction increases.

\section{SUMMARY AND DISCUSSION}

We have derived the distribution of tilt angles for coalescing BH--NS
binaries using very basic theoretical considerations for BH--NS formation.  
Our results show that the fraction of systems with tilt angles in excess
of 30$^{\circ}$ ranges from about 30\% to 80\% with a modest sensitivity
to the orbital separation of the BH--NS immediate progenitors and the kick
magnitude distribution.  Tilt angles in excess of
50$^{\circ}$--100$^{\circ}$ are expected for at most 70\% of the
coalescing BH--NS. Results obtained by \cite{A95}(1995) indicate that
aligned templates would be insufficient for more than 50\% of all binary
orientations if the spin tilt angle exceeds 30$^{\circ}$--40$^{\circ}$ and
for {\em all} binary orientations if the spin tilt angle exceeds
50--60$^{\circ}$. The implication is that the detection rate of BH--NS
coalescence events by ground-based laser interferometers (such as LIGO)
could be decreased by a factor up to $\simeq 4$, if waveform templates for
aligned spins are used in the data analysis.  It seems reasonable to
extend the database to precession-modified templates only for tilt angles
in the range 30$^{\circ}$ to 50$^{\circ}$, or at most to 100$^{\circ}$. We
note that the fraction of coalescing BH--NS with small spin tilt angles
increases (decreases) if kicks are preferentially directed perpendicular
(close) to the pre-SN orbital plane.

We can also constrain the binary properties of coalescing BH--NS. In
particular, we have shown that massive, hydrogen-rich immediate NS
progenitors are excluded and that a common-envelope phase is necessary. As
a result (i) circular pre-SN orbits and pre-SN spins aligned with the
orbital angular momentum axis are expected, and (ii) the pre-SN orbital
separations and the NS progenitor masses are restricted to $A_0\sim
10-100$\,R$_\odot$ and $M_0\sim 3-10$\,M$_\odot$. The expectation of
pre-SN alignment allows us to identify the tilt of the orbital planes
before and after the explosion with the BH spin tilt. The narrow ranges in
pre-SN parameters are primarily responsible for the robustness of our
results.

Previously, post-SN spin tilt angles have been studied in the context of
retrograde orbits in X-ray binaries and their possible connection to
long-term periodicities in these systems (\cite{BP95}1995), for BH--NS
binaries with a low-mass BH (3\,M$_\odot$) and a Roche-lobe filling NS, in
the context of precessing jets and their suggested association with
gamma-ray bursts (\cite{PZ99}1999), and for binary BH mergers
(\cite{PP99}1999). Based on the dependence of tilt angles on pre-SN binary
parameters and NS kick magnitude our results are in agreement with these
studies.

The inspiral waveform of a BH--NS binary is more strongly modified if the
spin of the BH (the more massive object in the system) is significantly
misaligned with respect to the orbital angular momentum axis.  The NS in
such systems is not expected to have been recycled in its lifetime (having
been formed after the BH). Therefore it would almost certainly be a slow
rotator at the time of the inspiral phase and the direction of its spin
would have no effect on the waveform.  If however, in addition to the BH
spin orientation, we were also interested in the spin orientation of the
NS, then we would need to make one additional assumption about the
physical origin of the NS spin. The generally accepted picture so far has
been that the rotation of the NS at birth is determined by the rotation of
the collapsing core, and hence the rotation of the NS progenitor. In this
case we would expect the NS spin to be aligned with its progenitor spin,
and hence the pre-SN orbital angular momentum axis. The angle $\omega$
then corresponds to both the BH and the NS spin tilt angle. However,
\cite{SP98}(1998) have argued recently that the origin of the NS spin may
be connected to the kick imparted to the NS at birth. Observational and
theoretical considerations (\cite{DRR00}2000; \cite{SP98}1998;
\cite{WKK00}2000) suggest that the kick timescale must be short enough
that the spin axis and kick direction are perpendicular (azimuthal
averaging about the spin axis is avoided).  Our analysis of the SN orbital
dynamics includes the kick direction.  Therefore, if the NS spin
orientation is of interest, it is possible to use this kick-spin
association to calculate the NS spin tilt angle distribution in BH--NS
binaries.

Spin--orbit coupling can in principle affect inspiral waveforms of
coalescing BH--BH binaries as well. However, as in the case of NS--NS
binaries, the effect is expected to be unimportant for equal-mass BH
binaries (\cite{A95}1995). It is only when the binary mass ratio 
 is small, as in a typical BH--NS system, that the modification of the
waveform can be significant, depending on the tilt angle. In Figure 9, we
plot the cumulative spin tilt angle distributions convolved with a
kick-magnitude distribution (Maxwellian with $\sigma=200$\,km\,s$^{-1}$)
and assuming isotropic kicks, for two different cases of BH--BH binaries:
one containing a 10\,M$_\odot$ and a 5\,$M_\odot$ BH and another with a
20\,M$_\odot$ and a 10\,M$_\odot$ BH. Comparison with our results for the
standard case (Figs.\ 5, 6)  indicates that BH--BH binaries tend to have
small tilt angles. More than $\sim 90$\% of the systems have angles
smaller than 30$^\circ$. Therefore, the effects of spin-orbit coupling on
BH--BH inspiral waveforms should be rather weak.

We note that in calculating the modifications of the inspiral waveforms in
the LIGO frequency band due to the spin-orbit misalignment, knowledge of
the spin tilt angles at the time the binary orbit enters the LIGO band is
required. The angles we derive in this paper characterize the tilts just
after the formation of the coalescing binary. One might worry that
gravitational radiation reaction effects could affect the spin orientation
as the binary approaches the final inspiral phases. It turns out that,
although the spin-orbit coupling is strong enough to modify the waveform
within the LIGO band, it is not strong enough to drive tilt angle
evolution on a fast timescale. \cite{R95}(1995) showed that the
misalignment angles at the time of the formation of the coalescing binary
do not change by more than one to a few per cent by the time the system
enters the inspiral phases of interest to ground-based laser
interferometers.

\acknowledgements

I am grateful to Bruce Allen and Ben Owen for bringing to my attention the
issue of spin orientation and detection of gravitational waves from close
BH--NS systems. I would also like to thank H.~Apostolatos, S.~Hughes,
E.~Flannagan, and A.~Wiseman for discussions on the tilt-angle evolution
due to gravitational radiation.  I acknowledge full support by the
Smithsonian Astrophysical Observatory in the form of a Harvard-Smithsonian
Center for Astrophysics Postdoctoral Fellowship.

\newpage

\begin{deluxetable}{cccccc}
\tablewidth{500pt}
\tablecaption{Limits on Isotropic Kick Magnitudes}
\tablehead{ \multicolumn{3}{c}{Model Parameters} &
$V_r$ & Minimum Kick & Maximum Kick \\
\colhead{$M_{\rm BH}$ (M$_\odot$)} & \colhead{$M_0$ (M$_\odot$)}
& \colhead{$A_0$ (R$_\odot$)} & (km\,s$^{-1}$) &
(km\,s$^{-1}$) & (km\,s$^{-1}$)}
\startdata
5 & 4 & 10 & 415 & 120 & 770 \\  
5 & 10 & 10 & 535 & 240 & 880 \\
5 & 4 & 50 & 185 & 135 & 300 \\
5 & 10 & 50 & 240 & 190 & 340 \\
\\
10 & 4 & 10 & 515 & 50 & 1035 \\
10 & 10 & 10 & 615 & 150 & 1125 \\
10 & 4 & 50 & 230 & 145 & 400 \\
10 & 10 & 50 & 275 & 195 & 435 \\
\\
20 & 4 & 10 & 675 & 0 & 1450 \\
20 & 10 & 10 & 755 & 0  & $>$1500  \\
20 & 4 & 50 & 300 & 160 & 545 \\  
20 & 10 & 50 & 340 & 195 & 575 
\enddata

\end{deluxetable}

\newpage

 \begin{figure}
 \plotfiddle{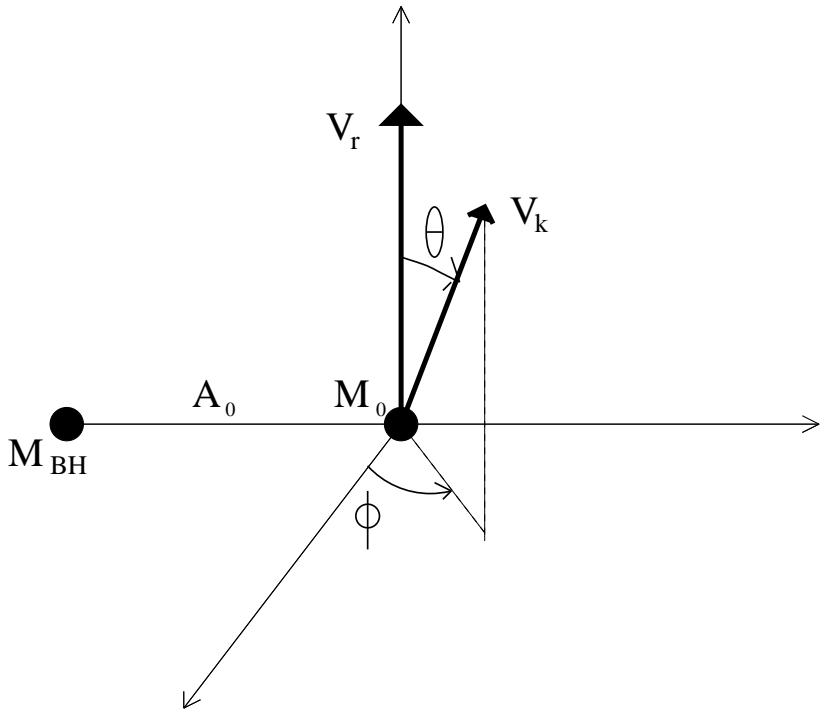}{4.0in}{-90}{70}{70}{-270}{450}
 \caption{Geometry of the binary system consisting of the black hole
$M_{\rm BH}$ and the NS progenitor $M_0$ at the time of the supernova
explosion. The pre-SN orbital plane coincides with the plane of the page.
$\vec{V_r}$ is the relative orbital velocity in the pre-SN orbit,
$\vec{V_k}$ is the kick imparted to the NS, and the angles $\theta$ and
$\phi$ define the kick direction.}
 \end{figure}

\newpage

 \begin{figure}
 \plotfiddle{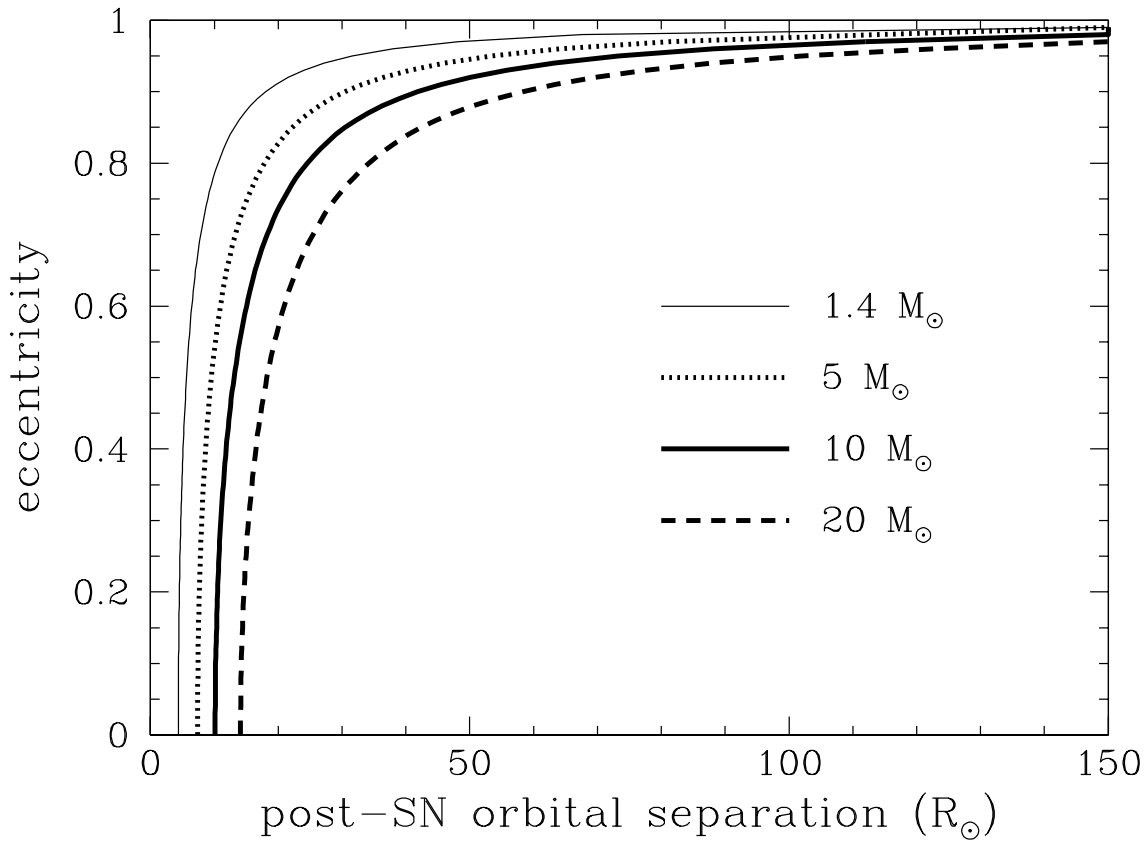}{4.0in}{-90}{70}{70}{-270}{450}
 \caption{Maximum post-SN orbital separation or minimum post-SN
eccentricity for coalescence to occur within 10$^{10}$\,yr, for NS
binaries with NS companions of different masses: 1.4\,M$_\odot$ (NS) and
5, 10, 20\,M$_\odot$ (BH).}
 \end{figure}

\newpage

 \begin{figure}
 \plotfiddle{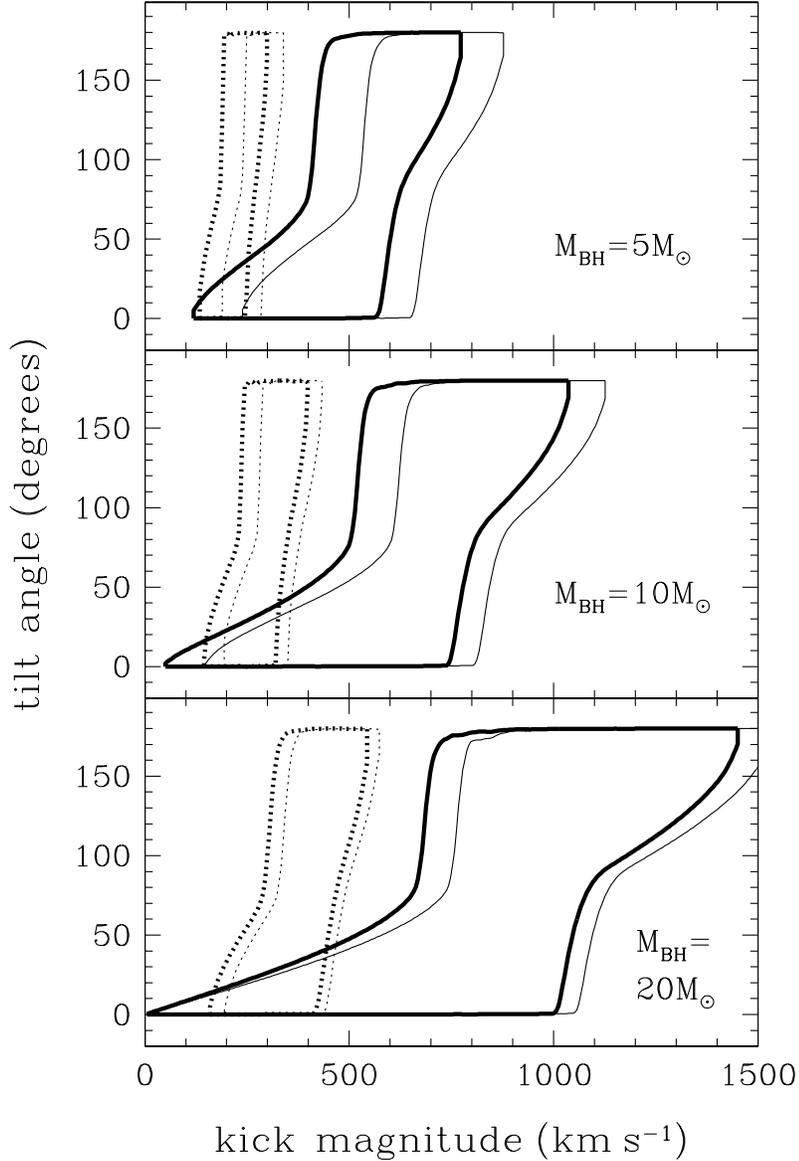}{5.in}{0}{60}{60}{-200}{-10}
 \caption{Upper and lower limits on the spin tilt angle in coalescing
BH--NS binaries as a function of an isotropic kick magnitude, for three
different BH masses and for different sets of pre-SN parameters (NS
progenitor mass and pre-SN orbital separation): 4\,M$_\odot$ (thick
lines), 10\,M$_\odot$ (thin lines), 10\,R$_\odot$ (solid lines), and
50\,R$_\odot$ (dotted lines).}
 \end{figure}

\newpage

 \begin{figure}
 \plotfiddle{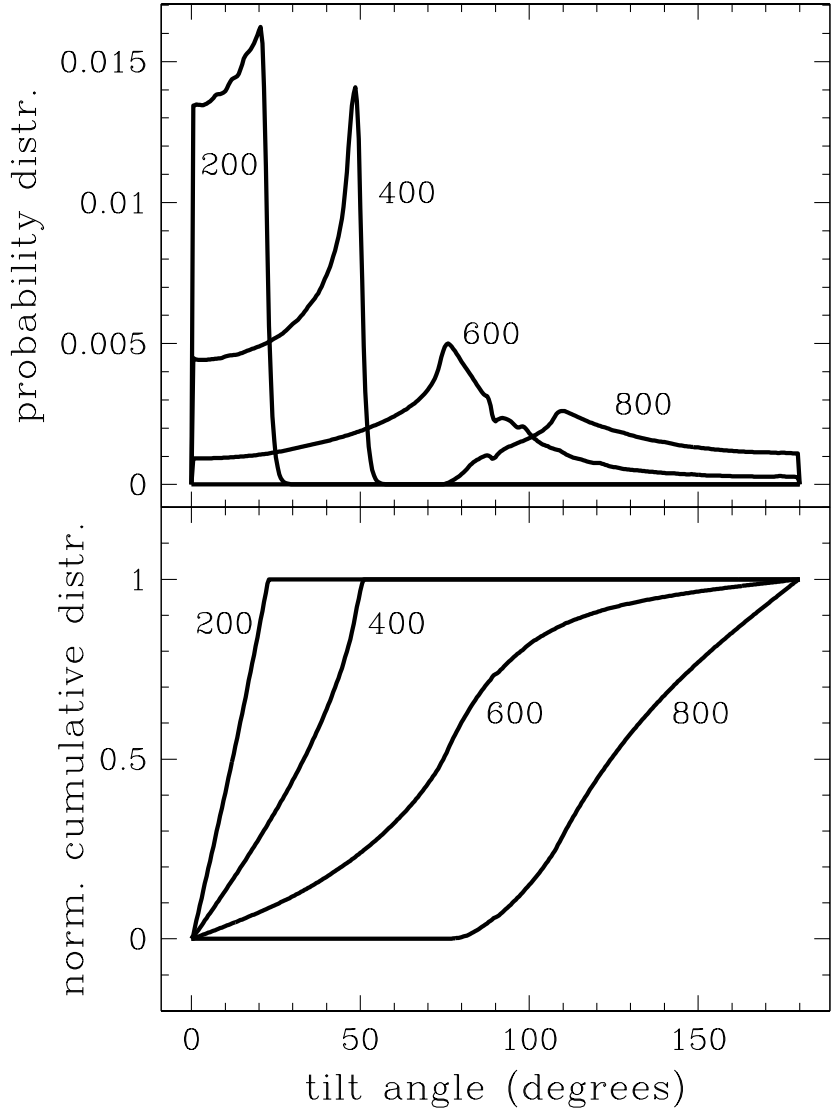}{5.in}{0}{60}{60}{-200}{-20}
 \caption{Probability distribution ({\em top}) and normalized (to unity)
cumulative distribution ({\em bottom}) of the spin tilt angle in
coalescing BH--NS binaries. Curves are plotted for $M_{\rm
BH}=10$\,M$_\odot$, $M_0=4$\,M$_\odot$, $A_0=10$\,R$_\odot$, and for four
different isotropic kick magnitudes, 200, 400, 600, 800\,km\,s$^{-1}$.
Note that the integrals over tilt angle of the distributions in the top
panel are equal to the fractions of post-SN systems that remain bound and
will coalesce within 10$^{10}$\,yr.}
 \end{figure}

\newpage

 \begin{figure}
 \plotfiddle{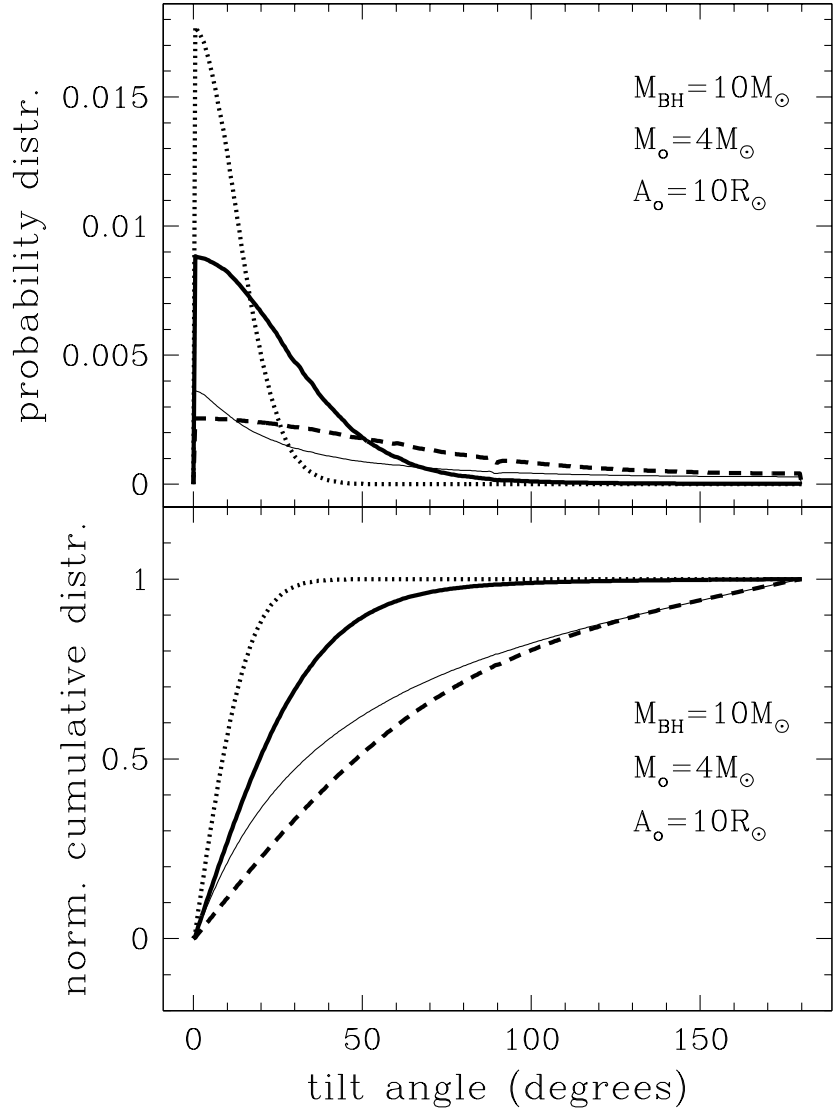}{5.in}{0}{60}{60}{-200}{-20}
 \caption{Probability distributions ({\em top}) and normalized cumulative 
distributions ({\em bottom}) of the spin tilt angle convolved with
four different kick magnitude distributions: Maxwellian with
$\sigma=100, 200, 400$\,km\,s$^{-1}$ (dotted, solid, dashed lines,
respectively) and a flat distribution in the range 0--1500\,km\,s$^{-1}$
(thin solid line). Normalization as in Figure 3.}
 \end{figure}

\newpage

 \begin{figure}
 \plotfiddle{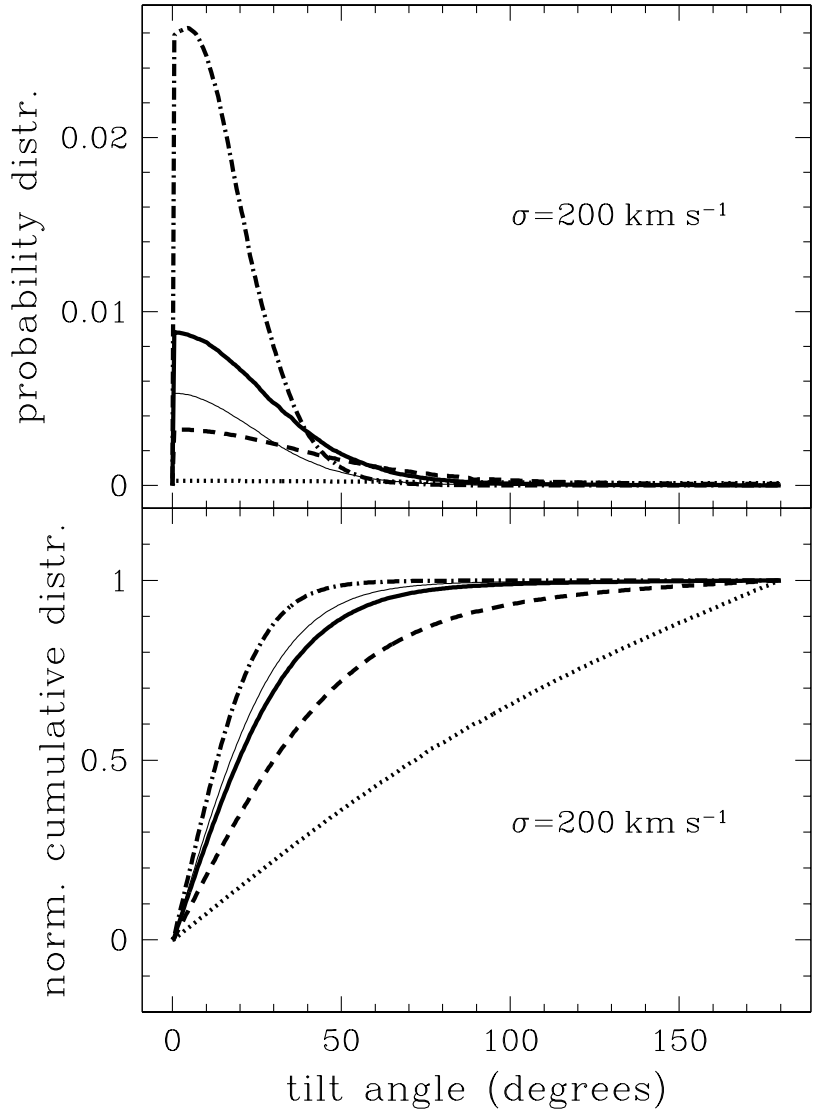}{5.in}{0}{60}{60}{-200}{-20}
 \caption{Probability distributions ({\em top}) and normalized cumulative
distributions ({\em bottom}) of the spin tilt angle convolved with a
Maxwellian kick magnitude distribution ($\sigma=200$\,km\,s$^{-1}$), for
five different sets of binary parameters: $M_0=4$\,M$_\odot$,
$A_0=10$\,R$_\odot$, and $M_{\rm BH}=5,10,20$\,M$_\odot$ (dashed, solid,
dot-dashed lines, respectively), $M_{\rm BH}=10$\,M$_\odot$,
$M_0=4$\,M$_\odot$, and $A_0=50$\,R$_\odot$ (dotted line), and $M_{\rm
BH}=10$\,M$_\odot$, $M_0=10$\,M$_\odot$, and $A_0=10$\,R$_\odot$ (thin
solid line). Normalization as in Figure 3.}
 \end{figure}

\newpage

\begin{figure}
 \plotfiddle{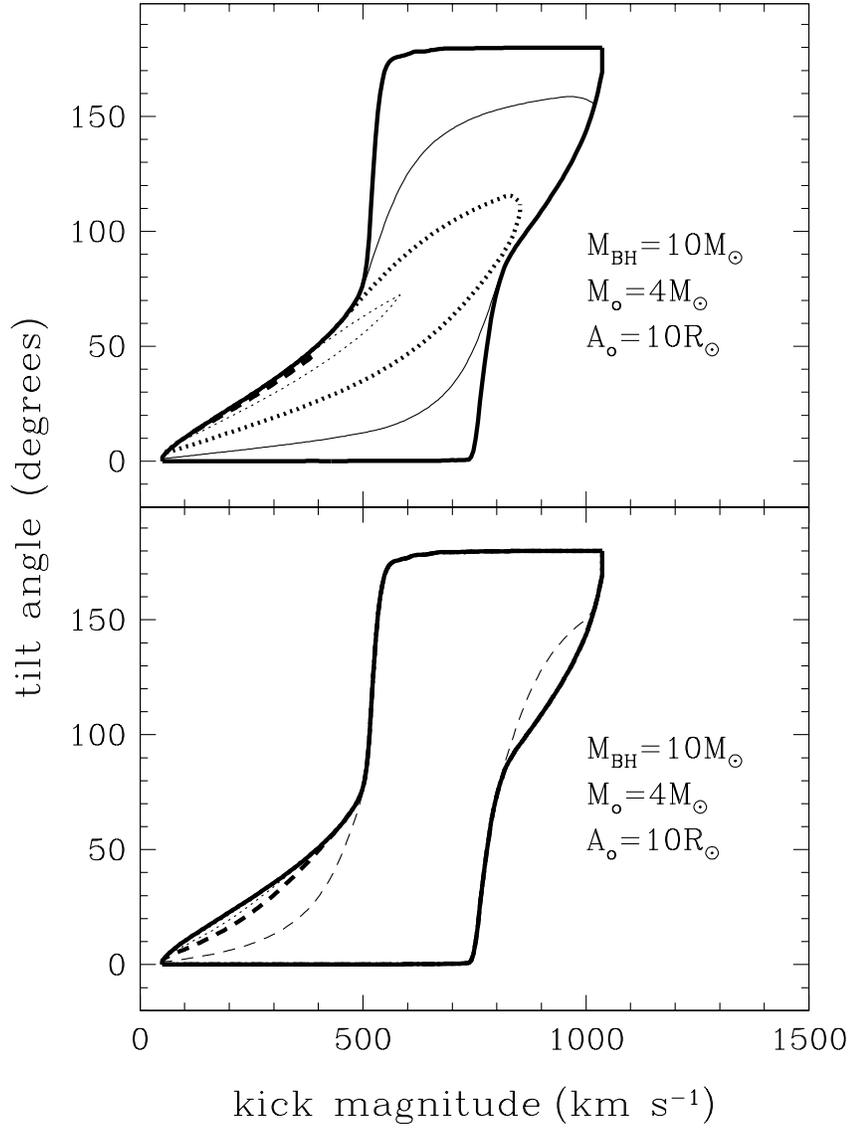}{5.in}{0}{60}{60}{-200}{-20}
 \caption{Upper and lower limits on the spin tilt angle in coalescing
BH--NS binaries as a function of the kick magnitude for
$M_{\rm BH}=10$\,M$_\odot$, $M_0=4$\,M$_\odot$,
$A_0=10$\,R$_\odot$, for polar kicks ({\em top}) and planar kicks
({\em bottom}), and for
$\theta_p=90^{\circ},80^{\circ},60^{\circ},40^{\circ},
30^{\circ},10^{\circ}$ (thick solid, thin solid, thick dotted, thin
dotted, thick dashed, and thin dashed lines, respectively).}
 \end{figure}

\newpage

\begin{figure}
 \plotfiddle{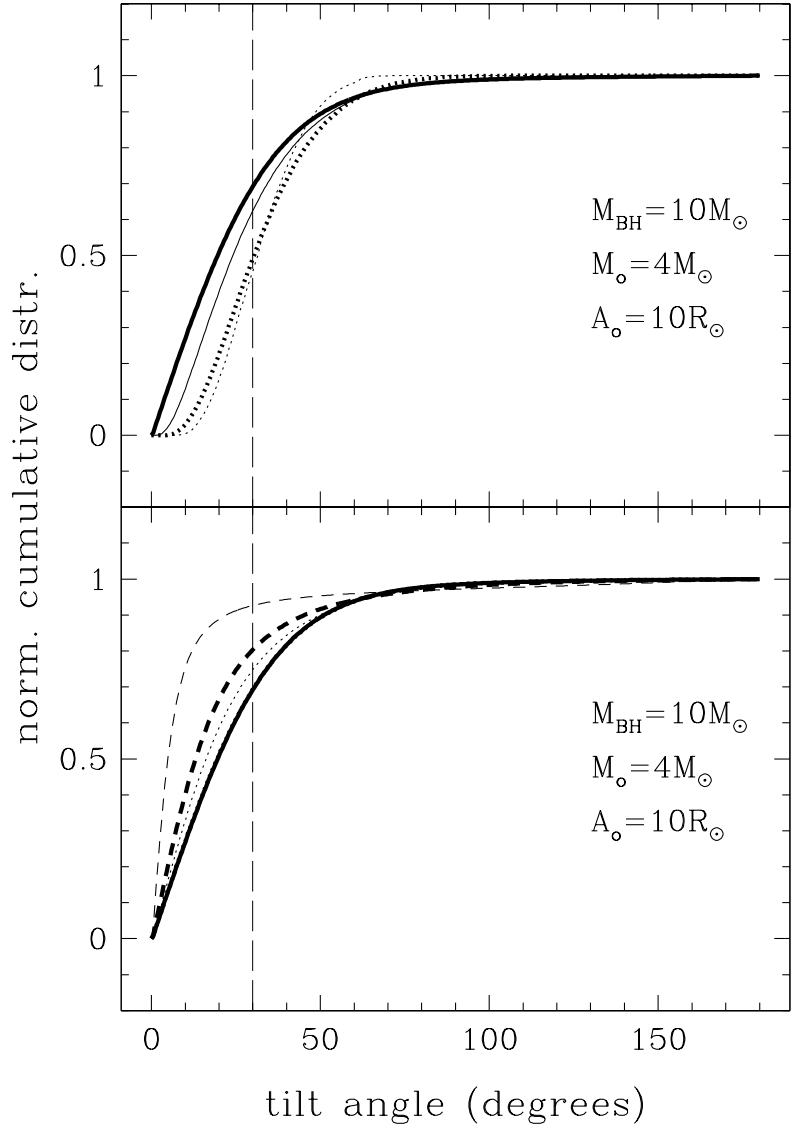}{5.in}{0}{60}{60}{-200}{-20}
 \caption{Normalized cumulative distributions of the spin tilt angle
convolved with a
Maxwellian kick magnitude distribution ($\sigma=200$\,km\,s$^{-1}$), for
$M_{\rm BH}=10$\,M$_\odot$, $M_0=4$\,M$_\odot$,
$A_0=10$\,R$_\odot$, for polar kicks (({\em top}) and planar kicks
({\em bottom}), and for
$\theta_p=90^{\circ},80^{\circ},60^{\circ},40^{\circ},
30^{\circ},10^{\circ}$ (line types same as in Figure 7).} 
 \end{figure}

\newpage

 \begin{figure}
 \plotfiddle{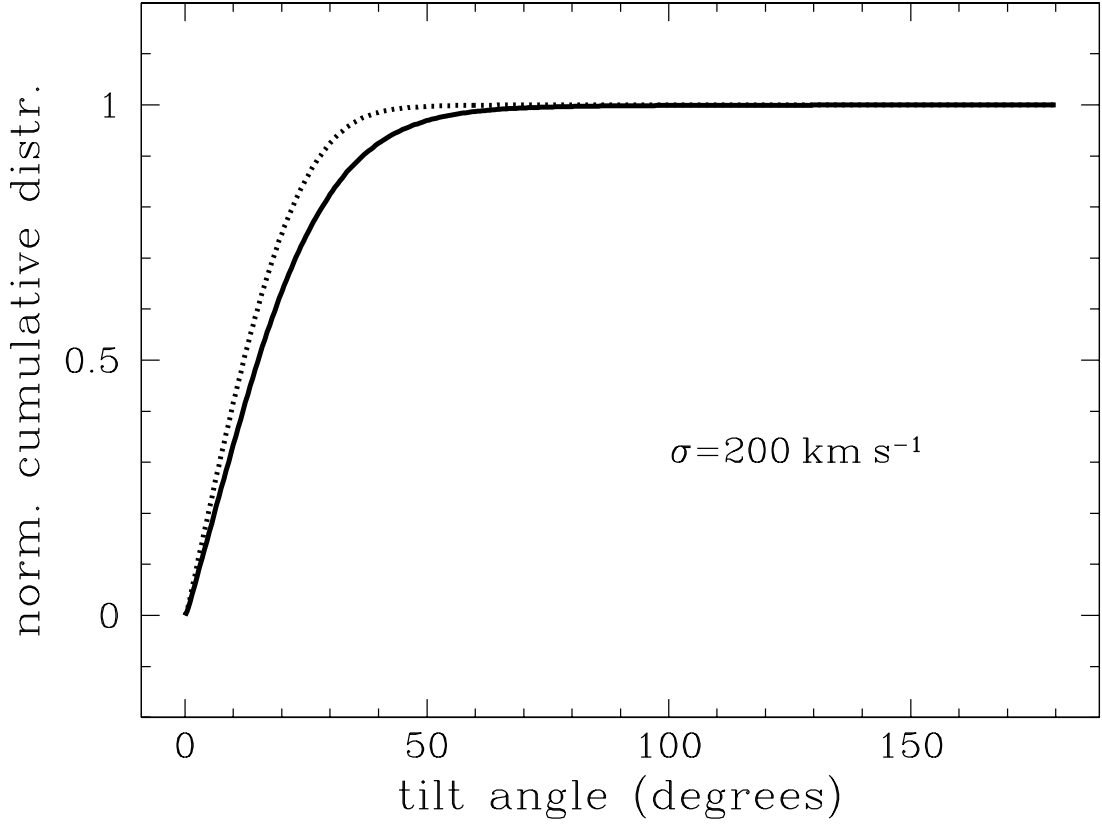}{4.0in}{-90}{70}{70}{-270}{450}
 \caption{Normalized (to unity) cumulative distributions of the spin tilt
angle convolved with a Maxwellian kick magnitude distribution
($\sigma=200$\,km\,s$^{-1}$), for BH--BH binaries: $M^1_{\rm
BH}=10$\,M$_\odot$, $M^2_{\rm BH}=5$\,M$_\odot$ (solid line), $M^1_{\rm
BH}=20$\,M$_\odot$, $M^2_{\rm BH}=10$\,M$_\odot$ (dotted line),
$M_0=10$\,M$_\odot$, and $A_0=10$\,R$_\odot$. Kicks are assumed to be
isotropic.}
 \end{figure}

\end{document}